\documentclass[doublecol]{epl2} 
\usepackage[latin1]{inputenc}
\usepackage{graphicx}
\usepackage{epsfig}
\usepackage{epstopdf}

\title{Nested Subgraphs of Complex Networks}
\shorttitle{Nested Subgraphs of Complex Networks} 

\author{Bernat Corominas-Murtra\inst{1} \and Jos\'e F. F. Mendes\inst{2} \and  Ricard V. Sol\'e\inst{1,3}}
\shortauthor{Corominas-Murtra \etal}

\institute{                    
  \inst{1} ICREA-Complex Systems
  Lab,  Universitat Pompeu  Fabra,  Dr.  Aiguader  80, 08003  Barcelona,
  Spain\\ 
  \inst{2}Departamento de F\'isica da Universidade de Aveiro, 3810-193 Aveiro, Portugal\\
  \inst{3}Santa  Fe Institute,  1399  Hyde Park  Road, New  Mexico
  87501, USA } 


\abstract{We analytically explore the  scaling properties of a general
class  of nested  subgraphs in  complex networks,  which  includes the
$K$-core and  the $K$-scaffold, among  others.  We name such  class of
subgraphs  $K$-nested subgraphs  due to  the fact  that  they generate
families  of   subgraphs  such  that   $...S_{K+1}({\cal  G})\subseteq
S_K({\cal  G})\subseteq S_{K-1}({\cal  G})...$.   Using the  so-called
{\em  configuration model}  it  is  shown that  any  family of  nested
subgraphs over a network with diverging second moment and finite first
moment  has  infinite  elements   (i.e.   lacking a percolation
threshold).   Moreover,  for  a  scale-free  network  with  the  above
properties,  we   show  that  any   nested  family  of   subgraphs  is
self-similar by  looking at  the degree distribution.   Both numerical
simulations and real data are analyzed and display good agreement with
our theoretical predictions. }

\begin{document}

\maketitle

\section{Introduction}
The internal  organization of most complex systems  displays some sort
of nestedness  associated to  some type of  hierarchical organization.
Such patterns  can be detected by using  appropriate theoretical tools
which  help us  understanding the  system's  structure in  terms of  a
network  \cite{Newman1, Farkas1,  Guimera,  Dorogovstevkcore1, Motifs,
scaffold, Bascompte}.  Furthermore,  the structure of such communities
can  provide us  valuable information  about invariant  properties and
potential universals.  In this work  we will define a general class of
network substructure which we  called $K-$nested subgraph.  Such class
of  subgraphs includes the  $K$-core, the  $K$-scaffold or  the random
deletion of  nodes.  But it  also includes any other  substructure you
can define, if it holds a small set of probabilistic restrictions.  We
develop a general, unified framework  that enables us to study generic
properties of such  $K-$nested subgraphs.  As we should  see, the most
common  class  of  real  networks, those  with  connectivity  patterns
following   a   power-law   distribution  $P(k)\propto   k^{-\alpha}$,
$2>\alpha>3$,  have  very   interesting  properties  when  looking  to
subgraph  nestedness.  In  this  context, theoretical  studies on  the
resilience  of both $K$-cores  \cite{Dorogovstevkcore1, CoresFernholz}
and $K$-scaffolds \cite{Shals,  scaffold} suggest that arbitrary large
scale-free  networks  contain  infinite, asymptotically  self-similar,
$K$-cores and $K$-scaffolds, indicating that such subgraphs are highly
robust against random deletion  of nodes.  Metaphorically, it has been
suggested that the  structure of complex nets is  similar to a Russian
doll \cite{Dorogovstevkcore1}.

These  results are  consistent with  the mounting  evidence indicating
that  scale-free  networks  exhibit  general  self-similar  properties
\cite{Nature,  Kim, Hamelin,  Dorogovstevkcore1, Guimera2}.   From the
physical  point of view,  the assymptotical  invariance of  the degree
distribution  of scale-free nets  under nesting  operations is  one of
their  most  salient  properties.    At  the  theoretical  level,  the
conservation    of   $P(k)$    the    degree   distribution    implies
self-similarity, as  far as most of  the properties of  a random graph
are determined by its degree distribution \cite{Bollobas}.  Of course,
real nets  are not exactly  random graphs, but such  approach revealed
surprisingly    adequate   to    study    real   systems\cite{Random}.
Furthermore,  self-similar properties  and  scaling laws  might be  an
indication   that  such   objects  are   organized   near  criticality
\cite{Gouyet, Mendes}.

In this  letter we generalize  previous approaches, showing  that {\em
any} nested family  of subgraphs of a given scale  free network has an
infinite percolation threshold i.e., there  is an infinite set of {\em
Russian dolls} for such networks.  Moreover, it can be shown that such
families  are  self-similar.   We  develop  such  concepts  under  the
framework of  the so-called {\em  configuration model} \cite{Besesky},
which  works   on  an  ensemble  of  arbitrarily   large,  sparse  and
uncorrelated graphs with specific properties.
\begin{figure}
  \includegraphics[width=7 cm]{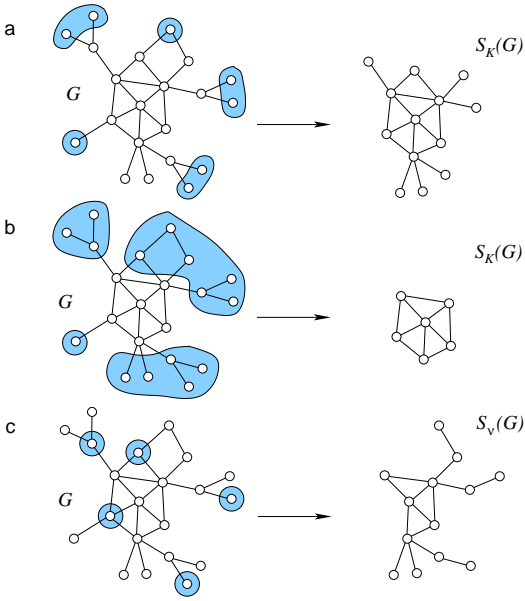}
  \caption{Some subgraphs  samples that enable  us to define  a nested
family of subgraphs.  In the original graph (left)
we shadowed the nodes that  disappear under the operation of $S_K$. In
the right-hand  side, we display  the giant component of  the obtained
graph, $S_K$.  We find the $K$-scaffold, ($K=3$) (a). The $K$-scaffold
is the subgraph obtained by  choosing all the nodes whose connectivity
is equal or higher than $K$  and all the nodes connected to them. Such
a subgraph enables to study the fundamental hub-connector structure of
the complex  networks. (b) The  $K$-core ($K=3$), the  largest induced
subgraph whose minimal  connectivity is equal to $K$.   (c) A subgraph
obtained by  randomly deleting a fraction ($\widehat{\nu}=5/21$) of nodes (commonly referred by
the literature as {\em random failures}.)}
  \label{Thresholds}
 \end{figure}
\begin{figure}
  \includegraphics[width=7.5 cm]{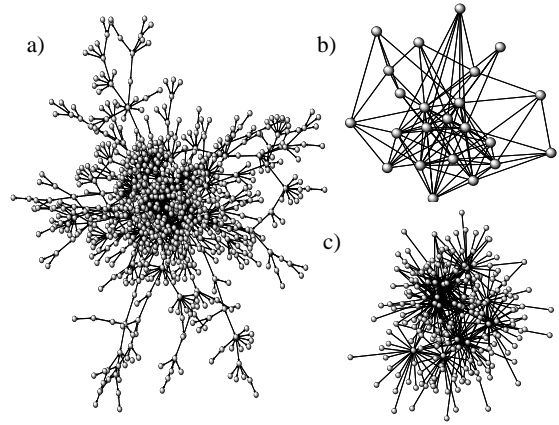}
  \caption{A  complex network with broad distribution of links (a)  
  and two  nested subgraphs:  (b) Its  $K$-core ($K=4$) and (c) the 
  corresponding  $K$-scaffold ($K=20$)}
  \label{Models}
 \end{figure}
The remaining of the paper is organized as follows: First, we formally
define the  concept of {\em $K$-nested}  subgraph and we  show how the
above  mentioned  examples hold  the  required  conditions.  Then,  we
derive the general percolation properties and the final, generic form,
of  an arbitrary  nested subgraph  of  a given net.  From the  developed
formalism, we apply our results to specific network topologies.

\section{Nested Subgraphs}
Formally,  a complex  network is  topologically described  by  a graph
${\cal   G}(V,  \Gamma)$   where  $V$   is  the   set  of   nodes  and
$\Gamma:V\rightarrow V$ the set of  edges connecting nodes of $V$.  If
$P(k)$ is the probability that  a randomly chosen node is connected to
$k$ other nodes, then
\begin{eqnarray}
\langle k  \rangle =\sum_k^{\infty} kP(k) \; \; \; \; 
\langle k^2 \rangle=\sum_k^{\infty} k^2 P(k) \nonumber
\end{eqnarray}
is the average connectivity of ${\cal G}$ and the second moment of the
distribution, respectively. 

We will say that $S(A,\Gamma_A)$ is  an induced subgraph of ${\cal
G}(V,  \Gamma)$ if $A  \subseteq V$  and $\Gamma_A  \subseteq \Gamma$,
being $\Gamma_A$ a mapping  $\Gamma_A:A \rightarrow A$.  We can define
many subgraphs from a given graph. Here we are interested in a special
set  of   subgraphs,  hereafter  {\em   $K$-nested  subgraphs},  which
includes,  as special  cases, the  family of  successive  $K$-cores or
$K$-scaffolds  and  the   so  called  $\widehat{\nu}-$deletion  graph,
obtained  by   deleting  a  fraction  $\widehat{\nu}$   of  nodes.   A
$K$-nested family of subgraphs ${\cal N}$ is a collection of subgraphs
of  a given  graph ${\cal  G}$, ${\cal  N}=\{S_1({\cal  G}), S_2({\cal
G}),...,S_i({\cal   G}),...\}$ such that:
\begin{equation}
...S_{K+1}({\cal  G})\subseteq  S_K({\cal  G})\subseteq  S_{K-1}({\cal G})...
\label{nest}
\end{equation}
\noindent
For every family  of $K$-nested subgraphs we associate  a {\em nesting
function} $\varphi_K(k)$, namely the probability for a randomly chosen
node with degree $k$ to belong to $S_K$. If $U\subseteq \mathbf{R}$ is
a set  that depends  on the nature  of the nesting,  $\varphi_K(k)$ is
such that:
\begin{equation}
\varphi_K(k):U\times  N\rightarrow [0,1] \nonumber
\end{equation}
It is easy  to see that, for  a function to be a  nesting function, it
has to fulfill the following logical conditions:
\begin{equation}
    (\varphi_K(k')>\varphi_K(k))\Rightarrow (k'>k)
\label{properties1}
\end{equation}
\begin{equation}
    (\varphi_{K'}(k)>\varphi_{K}(k))\Rightarrow (K'<K)
\label{properties2}
\end{equation}
\begin{equation}
    (\forall \varphi_K)(\exists \lambda_{S_K} \in (0,1])|(\lim_{k \to \infty}\varphi_K(k))= \lambda_{S_K})
  \label{properties3}
\end{equation}
where  $\lambda_{S_K}$ is  a scalar  whose  value will  depend on  the
explicit form of $S_K$.   In short, $\varphi_K(k)$ is a non-decreasing
function  on  $k$  (eq.   (\ref{properties1})) and  a  non  increasing
function on  $K$ (eq. (\ref{properties2})). Note that  such a function
implies that  all the nodes  satisfying the conditions are  taken into
account:  Our subgraphs are  maximal under  the conditions  imposed by
$\varphi_K$. Furthermore,  note that,  for a fixed  $K$, $\varphi_K(k)$
has   an    horizontal   asymptote   at   $\varphi_K(k)=\lambda_{S_K}$
(eq. (\ref{properties3})).  Thus:
\begin{equation}
\lim_{k \to \infty}(\varphi_K(k+1)- \varphi_K(k))=0
\label{lim1}
\end{equation}
\noindent
From    (\ref{properties1},    \ref{properties2},   \ref{properties3},
\ref{lim1}) we can  see that, for a fixed  $K$, and $0<\delta<1$ there
exist a $k^*$ such that:
\begin{equation}
(\forall k_i, k_j> k^*)\Rightarrow (||\varphi_K(k_i)-\varphi_K(k_j)||<\delta) \nonumber
\label{Cauchy}
\end{equation}
\noindent
and we can  conclude that the sequence $\{\varphi_K(k)\}=\varphi_K(1),
\varphi_K(2),..., \varphi_K(i),...$ is a Cauchy sequence. As we should
see, this property will be useful in the following sections.
Let us now explore some relevant nesting functions.


{\bf a)} {\em $K$-core subgraphs}. The $K$-core is the largest induced
subgraph whose minimal connectivity  is $K$.  Intuitively, it is clear
that a collection of $K$-cores from a given graph ${\cal G}$ defines a
nested family  of subgraphs.  Within  the configuration model,  we can
informally identify the probability for  a given node of ${\cal G}$ to
belong  to the giant  $K$-core with  the probability  to belong  to an
infinite   $(K-1)$-ary   subtree    of   ${\cal   G}$   \cite{Wormald,
CoresFernholz,  Dorogovstevkcore1}. 
Therefore, the probability for a  given node to belong to the $K$-core
equals  to the  probability of  belonging to  an  infinite $(K-1)$-ary
subtree.  Let  $R$ be the probability that  a given end of  an edge is
not  the root  of  an infinite  $(K-1)$-ary  subtree.  The  associated
nesting function for the $K$-core is $\varphi_K(k)=0, {\rm\;if\;} k<K$
and
\begin{equation}
\varphi_K(k)=\sum_{i=K}^{k}{k \choose i}R^{k-i}(1-R)^i \nonumber
\end{equation}
\noindent
otherwise. It is straightforward to check that such a function follows
(\ref{properties1}, \ref{properties2}, \ref{properties3}).

{\bf b)}{\em $K$-scaffold subgraphs} The $K$-scaffold of a given graph
is the subgraph  obtained by choosing all the  nodes whose $k\geq
K$ and the  nodes that, despite their connectivity  is $k<K$, they are
connected to a node $e'$ whose $k'\geq K$ \cite{Shals, scaffold}.
The nesting  function for the $K$-scaffold  is $\varphi_K(k)=1, {\;\rm
if\;}k\geq K$ and
\begin{equation}
\varphi_K(k)=1-\left(\sum_{k'<K}\frac{k'P(k')}{\langle k \rangle}\right)^k \nonumber
\end{equation}
\noindent
otherwise.    Note  that,   for  both   the  $K$-nested   families  of
$K$-scaffolds and $K$-cores, $\lambda_{S_K}=1$. A variety of subgraphs
can  be  defined  from  the  $K$-scaffold, such  as  the  {\em  naked}
$K$-scaffold  (a subgraph  obtained  by cutting  all  the nodes  whose
degree is $k=1$ in the $K$-scaffold).


{\bf c)}{\em Random deletion of  nodes}.- Suppose we delete a fraction
$\widehat{\nu}=1-\nu$ of  nodes from our graph. Such  an operation can
be also  formalized in  terms of nesting  functions.  For the  sake of
simplicity, if  we are performing a  random deletion of  a fraction of
nodes from ${\cal  G}$, we will indicate the  nesting function and the
subgraphs  as   $\varphi_{\nu}$  and  $S_{\nu}$,   respectively.   The
associated nesting function is, simply:
\begin{equation}
(\forall k)(\varphi_{\nu}(k)=\nu)
\label{nu}
\end{equation}

For  mathematical purposes, let  us introduce  an additional  class of
subgraphs, $S_{K\gamma}$, of a given subgraph $S_K$.  The main feature
of such subgraphs is that $S_{K\gamma} \subseteq S_K$.
We  name   such  subgraphs  {\em   minor  subgraphs}  of   $S_K$.   To
characterize  such subgraphs,  we say  that $\gamma_K(k)$  is  a minor
nesting function  of $\varphi_K(k)$ if  $(\gamma_K(k)<f_K(k))$ for all
$k$.  Given an arbitrary $\varphi_K(k)$,  we can build a minor nesting
function  as  follows:   Let  $k'$  be  the  minimum   $k$  such  that
$\varphi_K(k')\neq 0$ (it could be $k'=1$).  Then find an $\epsilon>0$
such that $\epsilon<\varphi_K(k')$.  Thus,
\begin{equation}
\gamma_K(k)=\left\{
\begin{array}{ll}
0\; {\rm if}\;k<k'\\
\epsilon  \; {\rm if} \; k\geq   k'
\end{array}
\right.
\label{minor}
\end{equation}
This  trivial way to  define a  minor subgraph  from a  given subgraph
$S_K$ is  enough, since both $\gamma_K(k)$  and $\varphi_K(k)$ verifie
(\ref{properties1},\ref{properties2},\ref{properties3}).  Moreover, it
is  clear that\footnote{Let  us suppose  a  graph ${\cal  G}$ and  two
subgraphs of  it, $S_{\nu}, S_{\nu'}$, obtained by  deleting at random
$\widehat{\nu}=1-\nu$  and  $\widehat{\nu}'=1-\nu'$, with  $\nu'>\nu$.
Clearly, we cannot  conclude that $S_{\nu}$ is an  induced subgraph of
$S_{\nu'}$.  But it is true  that the properties of $S_{\nu}$ will be,
with  high probability,  the properties  of some  induced  subgraph of
$S_{\nu'}$  obtained by  deleting at  random $\widehat{\nu}$  nodes of
${\cal G}$.  Furthermore, it can be shown that the probability to find
a diverging  value decays  exponentially with the  size of  the system
-Recall   that   we  are   working   with   an  ensemble   formalism.}
$(S_{K\gamma} \subseteq S_K)$ for all $K$.

\section{ Percolation  of nested  subgraphs}
Previous  to determining  the specific  statistical properties  of the
obtained subgraphs,  we are interested  in knowing whether there  is a
giant component in $S_K$, i.e., if the operation of nesting breaks (or
not)  the initial  graph ${\cal  G}$ into  many small  components.  We
consider first the general problem.

Let  us define the  generating functions  for an  arbitrary $K$-nested
subgraph with an associated nesting function $\varphi_K(k)$ defined on
${\cal G}$ with arbitrary (but smooth) degree distribution $P(k)$.
\begin{eqnarray}
F_0(z)=\sum_{k}^{\infty}P(k)\varphi_K(k)z^k \\
F_1(z)=\frac{1}{\langle k \rangle}\sum_k^{\infty}kP(k)\varphi_K(k)z^{k-1}
\end{eqnarray}
The averages  -i.e., the values  at $z=1$ of  eqs.  (5) and  (6)- are,
respectively,  $\mu\equiv F_0(1)$ and  $\omega  \equiv F_1(1)$.   Here,
$\mu$ is the  fraction of nodes from ${\cal G}$  that belong to $S_K$.
Similarly, $\omega$ is the relation  among $\langle k \rangle$ and the
average number of nodes from  $V$ reachable after computing the nested
subgraph.  The  generating function for the size  of components -other
than the giant component- which  can be reached from a randomly chosen
node is:
\begin{equation}
H_1(z)=1-\omega+zF_1(H_1(z))\nonumber
\label{H1}
\end{equation}
\noindent
and the generating  function for the size of the  component to which a
randomly chosen node belongs to is \cite{Callaway, Random}:
\begin{equation}
H_0(z)=1-\mu+zF_0(H_1(z))\nonumber
\label{H0}
\end{equation}
\noindent
thus, the average  component size other than the  giant component is:
\begin{equation}
\langle s \rangle=H'_0(1)=\mu+F'_0(1) H'_1(1)\nonumber
\label{s}
\end{equation}
If we compute the derivative, it  is straightforward  to see  that it  leads to  a  singularity when
$F'_1(1)=1$. Thus, if
$F'_1(1)= \frac{1}{\langle k \rangle}\sum_kk(k-1)\varphi_K(k)P(k)$
, to ensure the presence of  a giant $S_K$, the following inequality has
to hold:
\begin{equation}
\sum_k k(k-2)P(k)>\sum_k k(k-1)\widehat{\varphi}_K(k)P(k)\nonumber
\label{Percol1}
\end{equation}
\noindent
Where  $\widehat{\varphi}_K(k)=1-\varphi_K(k)$.  This  can be  seen as
the   natural   extension   of   the   Molloy   and   Reed   criterion
\cite{MolloyReed}  for  any  nested  subgraph $S_K$,  with  associated
nesting function $\varphi_K(k)$.  A  more compact expression of such a
criterion is:
\begin{equation}
  \sum_k k^2 \varphi_K(k) P(k)-(1+\omega)\langle k \rangle > 0
  \label{Percol2}
\end{equation}
\begin{figure}
  \includegraphics[width=8.2 cm]{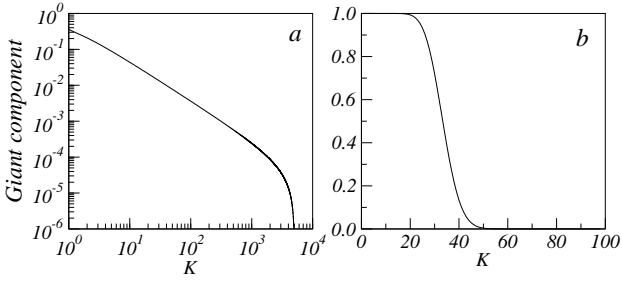}
  \caption{The  simplest  family  of  nested  subgraphs,  obtained  by
removing   all   nodes   whose   connectivity  is   less   than   $K$:
$\varphi_K(k)=\Theta(K,k)$,  where $\Theta(K,k)=1$  iff $k\geq  K$ and
$0$ otherwise.   (a) Numerical  computation of the  size of  the giant
component $p_{\infty}=1-H_0(1)=\mu-F_0(u)$ where $u$ is the first, non
trivial        solution       of        $u=1-\omega+F_1(u)$,       for
$\varphi_K(k)=\Theta(K,k)$.   This curve  corresponds to  a scale-free
network with $\alpha \approx  2.15$.  No specific scale is identified.
The sharp  decay for  the large  $K$ values can  be attributed  to the
finite   size  of  the   system  (In   this  simulation,   we  assumed
$k_{max}\approx  5000$ ).  (b)  The same  computation over  an Erd\"os
R\'enyi  graph   with  $\langle  k  \rangle  =30$   displays  a  clear
characteristic   scale  where  the   giant  component   is  completely
eliminated.}
\label{Heavyside}
\end{figure}

\section{Degree distribution of $S_K$}
The  next step is  to compute  the degree  distribution of  the nested
subgraphs,  $P_{S_K}(k)$.  The  key  question is  finding the  average
number  of nodes  a  given node  will  reach, if  it  survived to  the
computation of  $S_K$.  Taking  into account the  set of all  nodes of
${\cal G}$,  the average connectivity  will decrease a  factor $\omega
\equiv  F_1(1)=1/\langle  k  \rangle \times\sum_k  k\varphi_K(k)P(k)$.
Clearly, the probability for a {\em surviving} node whith connectivity
$k$  in  ${\cal G}$  to  display  connectivity  $k'\leq k$  in  $S_K$,
${\mathbf P}(k\rightarrow k')$, is:
\begin{equation}
{\mathbf P}(k\rightarrow k')={k \choose k'}\omega^{k'}(1-\omega)^{k-k'}
\end{equation}
And,  in absence  of correlations,  a node  whith connectivity  $k$ in
${\cal G}$  now will {\em  survive} with a  probability $\varphi_K(k)$
and it will be connected, on average, to $\omega k$ nodes.  If we take
into account all the possible contributions of the nodes of ${\cal G}$
to the abundance of nodes with certain degree $k$ in $S_K$, we have:
\begin{equation}
P_{S_K}(k)=\frac{1}{\mu}\sum_{i\geq
k}^{\infty}\varphi_K(i){i \choose k}\omega^{k}(1-\omega)^{i-k}P(i)
\end{equation}
Where $P_{S_K}(k)$  is the  probability to find  a node of  degree $k$
after the computation of  $S_K$.  Note that the factor $\frac{1}{\mu}$
normalizes      $P_{S_K}(k)$.      Clearly,      if      we     define
$\delta(\omega,\lambda_{S_K})$ as:
\begin{eqnarray}
\delta(\omega,\lambda_{S_K})\equiv \frac{1}{\mu}\sum_{i\geq
k}^{\infty}(\lambda_{S_K} -\varphi_K(i)){i \choose k}\omega^{k}(1-\omega)^{i-k}P(i)
\nonumber
\end{eqnarray}
We can rewrite $P_{S_K}$ as:
\begin{equation}
P_{S_K}(k)=\frac{\lambda_{S_K}}{\mu}\sum_{i\geq k}^{\infty}{i \choose k}\omega^{k}(1-\omega)^{i-k}P(i)-
\delta(\omega,\lambda_{S_K})\nonumber
\label{P}
\end{equation}
But note that, due to relation (\ref{Cauchy}), for large $k$'s:
\begin{equation}
\frac{\lambda_{S_K}}{\mu}\sum_{i\geq k}^{\infty}{i \choose k}\omega^{k}(1-\omega)^{i-k}P(i)\gg 
\delta(\omega,\lambda_{S_K})
\nonumber
\end{equation}
Thus $P_{S_K}$ is reduced to:
\begin{equation}
P_{S_K}(k) \approx\frac{\lambda_{S_K}}{\mu}\sum_{i\geq k}^{\infty}{i \choose k}
\omega^{k}(1-\omega)^{i-k}P(i)
\label{DegBin}
\end{equation}
Let us rewrite equation  (\ref{DegBin}) in order to extract analytical
results. If the first generating function of the degree distribution of
${\cal G}$, without taking into account the nesting operation, is:
\begin{equation}
G_0(z)=\sum_k^{\infty}P(k)z^k \nonumber
\end{equation}
It is straightforward that:
\begin{equation}
\frac{d^k}{dx^z}G_0(z)=\sum_{i\geq k}\frac{i!}{(i-k)!}P(k)z^{i-k}\nonumber
\end{equation}
Thus, we  can rewrite the degree distribution  (\ref{DegBin}) in terms
of the derivatives of $G_0(z)$:
\begin{eqnarray}
P_{S_K}(k)\approx
\frac{\lambda_{S_K}}{\mu}\frac{\omega^k}{k!}\left. \frac{d^k}{dz^k}G_0(z)\right|_{z=1-\omega}
\label{PS}
\end{eqnarray}
In the following, we will  apply our results to standard topologies of
network theory:  The Erd\"os R\'enyi graphs and  the Power-law graphs.

\section{Erd\"os R\'enyi Graphs}.   
In the Erd\"os R\'enyi (E-R) graph,
\begin{equation}
P(k)=\frac{\langle  k  \rangle^ke^{\langle   k  \rangle}  }{k!}\nonumber  
\end{equation}
and $\langle  k^2 \rangle=\langle  k \rangle^2$.  To  study specifical
percolation  preoperties,  we  need  to  know the  specific  shape  of
$\varphi_K(k)$. In  (fig.\ref{Heavyside}-b)) we approached numerically
the  size of  the giant  component  in an  E-R graph  where a  nesting
successive  nesting  operation  is  performed. A  clear  threshold  is
observed,  displaying  a  critical  point where  the  giant  connected
component   is   completely   eliminated.    The   special   case   of
$\varphi_K(k)=\nu$ recovers  the well-known percolation  condition for
E-R    graphs    under    random    damage,   $\langle    k    \rangle
>(1+\nu)/\nu$.  The predictions  for the degree distribution are
more general and accurate.  Indeed, the expression for $G_0(z)$ in E-R
graphs is  $G^{ER}_0(z)=e^{\langle k \rangle(z-1)}$.  Thus,  if, as we
defined above, $\mu\equiv F_0(1)$ :
\begin{eqnarray}
P^{ER}_{S_K}(k)
\approx \frac{\lambda_{S_K}}{\mu}\frac{\langle\omega    k \rangle^ke^{\langle \omega k \rangle} }{k!}
\end{eqnarray}
This implies that, for large  $k's$, the nesting operation over an E-R
graph results  in an E-R graph  but with a  factor $\omega$ correcting
the  mean  value,  whose  value  goes  from  $\langle  k  \rangle  \to
\omega\langle k \rangle$.  
\begin{figure}
  \includegraphics[width=8.6 cm]{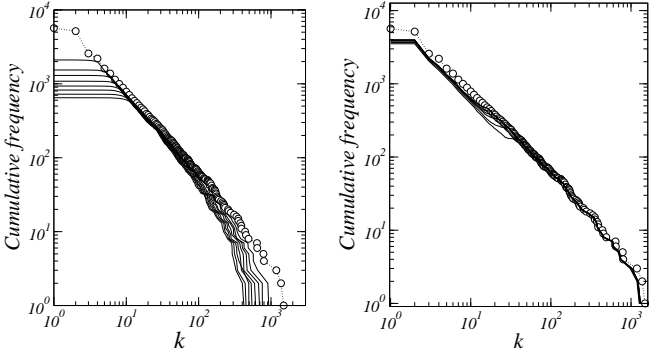}
  \caption{Analyzing the web obtained  from the O.  Wilde's novel {\em
  The portrait of Dorian Gray}. The network was built up by tracing an
  arc between two  adjacent words, if they appear  one after the other
  within the same sentence.  The obtained graph has $N=5696$ nodes and
  displays a  scale-free distribution $P(k)\propto  k^{-\alpha}$ (grey
  circles),  with  an   exponential  cut-off  at  high  connectivities
  ($k>1000$).   In this graph,  $\alpha \approx  2.15$ and  $\langle k
  \rangle \approx 8.814$  .  We plot the cumulative  frequency for the
  $K$-cores, $4\geq K\geq 11$ (left).  Despite the strong connectivity
  requeriments imposed  for the $K$-core, the  distribution behaves as
  an statistical invariant.  The same is observed with successive {\em
  naked} $K-$scaffold subgraphs, $K=14,16,18,20,22,30,40$ (right). The
  naked $K-$scaffold  subgraph is  obtained from the  $K$-scaffold but
  deleting all  the nodes  with $k<K$ that  are connected only  to one
  node with $k'\geq K$.}
  \label{Distrib}
 \end{figure}

\section{Scale-free  nets} 
Let us assume a scale-free network with
\begin{equation}
P(k)\propto k^{-\alpha}
\end{equation}
\noindent
with  scaling  exponent  $2<\alpha<3$.   We  will show  that,  at  the
thermodynamic  limit,  {\em  any}  family of  subgraphs  has  infinite
subgraphs.   This  has   been  shown   separately  for   the  $K$-core
\cite{CoresFernholz,    Dorogovstevkcore1}   and    the   $K$-scaffold
\cite{scaffold}.  One of the main characteristics of such nets is that
$\langle k^2  \rangle \to \infty$,  and that $\langle k  \rangle$ does
not diverge with network size.

What  we  should  prove  is  that, under  these  conditions,  relation
(\ref{Percol2}) holds {\em  for all} $K$'s.  In other  words, there is
no   characteristic   scale   for   the  substructure   generated   by
$\varphi_K(k)$.  Indeed, our subgraphs need to fulfill the inequality:
\begin{equation}
\sum_k k^2 \varphi_K(k)P(k)-(1+\omega)\langle k \rangle>0\nonumber
\end{equation}
But  we  cannot  work  directly  with an  arbitrary  nesting  function
$\varphi_K$. Thus, to prove the  above claim, we build a minor nesting
function   $\gamma_K(k)$  of   our  $\varphi_K(k)$,   as   defined  in
(\ref{minor}),   assuming  $k'$   as  the   smallest  $k$   such  that
$\varphi_K(k)>0$.   Thus,  if $\omega_{\gamma}\equiv  F^{\gamma}_1(1)$
has the form:
\begin{equation}
\omega_{\gamma}=\epsilon \left(1- \sum_{k<k'}\frac{kP(k)}{\langle k\rangle}\right)\equiv
\epsilon'\nonumber
\end{equation}
The corresponding percolation  condition for $S_{K\gamma}$ is, thus:
\begin{eqnarray}
\epsilon\sum_{k\geq k'} k^2P(k)-(1+\epsilon')\langle k \rangle>0\nonumber
\label{phi}
\end{eqnarray}
But   since   $\langle   k^2   \rangle$   diverges,   we   will   have
$\epsilon\sum_{k\geq   k'}k^2P(k)\rightarrow  \infty$   and  condition
(\ref{Percol2})  always holds,  provided that  $\langle k  \rangle$ is
finite.  This  implies that percolation  of any nested subgraph  of an
arbitrary  large   scale-free  network   is  guaranteed,  as   far  as
$S_{K_{\gamma}}\subseteq    S_K$.     Numerical    simulations    (see
(fig\ref{Heavyside}-a)) of the size  of the giant component display no
critical scale for the  emergence (elimination) of the Giant connected
component.

The above mathematical machinery will  lead us to demonstrate that our
families   of   nested   subgraphs   exhibit  invariance   in   degree
distribution.     If    we    put   the    distribution    $P(k)={\cal
C}^{-1}k^{-\alpha}$,  (${\cal C}=\zeta(\alpha)$),  equation (\ref{PS})
becomes to:
\begin{eqnarray}
P_{S_K}(k)\approx \lambda_{S_K}\frac{\omega^k}{k!}\left. \frac{d^k}{dz^k}G^{SF}_0(z)\right|_{z=1-\omega}
\end{eqnarray}
Thus  the  problem  lies  on  finding the  {\em  k-th}  derivative  of
$G^{SF}_0(z)$.  The computation is  slightly more complex than the E-R
graphs, and involves some approaches. First, we compute the generating
function for  an scale-free net  $P(k)={\cal C}^{-1}k^{-\alpha}$ whose
exponent lies between $2$ and $3$, $G_0^{SF}(z)$:
\begin{eqnarray}
G_0^{SF}(z)
 ={\cal C}^{-1}\mathbf{Li}_{\alpha}(z)\;\;\;\;\;\;\;\;\;\;\;\;\;\;\;\;\;\;\;\;\nonumber \\
\;\;\;\;\;\;\;\;\;\;\;\;\;\;\;\;\;\;\;\;
={\cal C}^{-1}\frac{z}{\Gamma(\alpha)}\int^{\infty}_0 
dt\frac{t^{\alpha -1}}{e^t-z}\nonumber
\end{eqnarray}
Where   $\mathbf{Li}_{\alpha}(z)=\sum_k^{\infty}\frac{z^k}{k^{\alpha}}$
is the  polylogarithm function and, to  obtain the last  step, we used
its  integral   form.   But,  actually,  we  are   interested  in  the
derivatives  of $G_0^{SF}(z)$.   If we  assume $z\to  1^-$  the $k-th$
derivative of $G_0^{SF}(z)$ can be approached by:
\begin{eqnarray}
\frac{d^k}{dz^k}G_0^{SF}(z) 
\approx {\cal C}^{-1}\frac{k!}{\Gamma(\alpha)}\int^{\infty}_0 
dt\frac{t^{\alpha -1}}{(e^t-z)^{k+1}}\nonumber \\
\approx {\cal C}^{-1}\frac{k!}{\Gamma(\alpha)}\int^{\infty}_0 
dt\frac{t^{\alpha -1}}{(t+\tau)^{k+1}}\nonumber \\
={\cal C}^{-1}\frac{k!\tau^{\alpha-1-k}}{\Gamma(\alpha)}\int^{\infty}_0 dy 
\frac{y^{\alpha -1}}{(y+1)^{k+1}}\nonumber
\end{eqnarray}
Where, in the  first approach, we used the fact that,  if $z\to 1$, we
are near a singularity when $t\to 0$.  Thus, the dominant terms of the
sum  will  be  those  close  to  $t=0$. This  enables  us  to  rewrite
$e^t\approx  1+t+{\cal  O}(t^2)$.   In  the  last step,  we  made  the
coordinate  change $\tau=1-z$  and, then,  $t=y\tau$.  If  we evaluate
such an expression at $z=1-\omega$, with $\omega$ small enough:
\begin{eqnarray}
\left.\frac{d^k}{dz^k}G_0^{SF}(z)\right|_{z=1-\omega}\approx {\cal C}^{-1}\frac{k!\omega^{\alpha-1-k}}{\Gamma(\alpha)}J_{k+1, \alpha+1}\nonumber
\end{eqnarray}
Where $J_{k+1,\alpha+1}$ is defined as:
\begin{eqnarray}
J_{k+1,\alpha+1}\equiv \int^{\infty}_0 dy \frac{y^{\alpha -1}}{(y+1)^{k+1}}
=\frac{\Gamma(\alpha)\Gamma(k-\alpha+1)}{k!(k-\alpha+2)} \nonumber
\end{eqnarray}
If  we check the  behavior of $J_{k+1, \alpha+1}$   for large
$k$'s, we see that:
\begin{equation}
J_{k+1, \alpha+1} \approx \frac{\Gamma(\alpha)}{k^{\alpha}}
\end{equation}
Thus,  if  we introduce  the  above  results  into the  definition  of
$P_{S_K}$:
\begin{eqnarray}
P_{S_K}(k)\approx \frac{\lambda_{S_K}}{\mu}\frac{\omega^k}{k!}\left.\frac{d^k}{dz^k}G_0^{SF}(z)
\right|_{z=1-\omega} \nonumber \\
={\cal C}^{-1}\frac{\lambda_{S_K}}{\mu}\omega^{\alpha -1}{k^{-\alpha}}
\end{eqnarray}
Which  can  be rewritten  in  the  standard  form when  describing  of
self-similar objects:
\begin{eqnarray}
P_{S_K}(k)\approx\rho^{-\alpha}P(k)=P(\rho k)
\label{Self}
\end{eqnarray}
Where $\rho$ is a constant  that, interestingly, depends both with the
scaling  exponent $\alpha$  and the  nature  of the  nesting, namely:
\begin{equation}
\rho  =\left(\frac{\mu}{\lambda_{S_K}\omega^{(\alpha-1)}}\right)^{\frac{1}{\alpha}}
\end{equation}

\section{Discussion}
Many  interacting  systems  found   in  nature  display  a  scale-free
topology,  $P(k)\propto  k^{-\alpha}$, with  $2<\alpha<  3$.  In  this
letter we have  shown that the assumptions of  the configuration model
are enough to explain many  of the scaling and self-similar properties
of  the  observed nested  subgraphs  nets.   The resulting  prediction
(\ref{Self})  reveals that,  under no  correlations, we  should expect
invariance in degree distributions of nested subgraphs to occur.  This
is  what  we   observe  in  the  analysis  of   real  nets  (see  fig.
(\ref{Distrib})).  Indeed, in the  analysis of the degree frequency we
see that, despite the finite  size of our system, the degree frequency
acts  as an  invariant, only  modulated by  an scaling  factor.  These
results contrast  with previous work on sampled  subnets obtained from
scale-free graphs \cite{May}.  Although is true that arbitrary subsets
of  nodes  might  not  display  invariance,  our  families  of  nested
subgraphs are defined  in such a way that our  results are expected to
hold. Further work should address the impact of the self-similarity in
the  functional aspects  of the  net, as  well as  a broader  study of
nested subgraphs involving different types of real networks.

\acknowledgments The authors thank  the members of the Complex Systems
Lab  for useful  comments.  This  work  has been  supported by  grants
FIS2004-0542,  IST-FET  ECAGENTS, project  of  the European  Community
founded under EU  R\&D contract 01194, FIS2004-05422 and  by the Santa
Fe Institute.

\end{document}